\documentclass[aps,twocolumn,superscriptaddress,preprintnumbers,floats,floatfix]{revtex4}
\usepackage[colorlinks, citecolor=blue,anchorcolor=red,menucolor=red, linkcolor=red,filecolor=red,urlcolor=blue,frenchlinks=red,hypertexnames=false]{hyperref}
\usepackage{amsfonts}
\usepackage{amsmath}
\usepackage{amssymb}
\usepackage{CJKutf8}
\usepackage{color}
\usepackage{comment}
\usepackage{epsfig}
\usepackage{epstopdf}
\usepackage{float}
\usepackage{graphicx,booktabs}
\usepackage{indentfirst}
\usepackage{longtable,lscape}
\usepackage{mathrsfs}
\usepackage{mathtools}
\usepackage{morefloats}
\usepackage{pifont}
\usepackage[T1]{fontenc}
\usepackage{txfonts}
\usepackage{multirow}

\begin{document}

\title{Where are the higher excited $B$ and $B_s$ mesons with coupled-channel effects?}

\author{Ru-Hui Ni}
\email{niruhui@ucas.ac.cn}
\affiliation{School of Physical Sciences, University of Chinese Academy of Sciences, Beijing 100049, China}
\author{Jia-Jun Wu}
\email{wujiajun@ucas.ac.cn}
\affiliation{School of Physical Sciences, University of Chinese Academy of Sciences, Beijing 100049, China}
\affiliation{Southern Center for Nuclear-Science Theory (SCNT), Institute of Modern Physics, Chinese Academy of Sciences, Huizhou 516000, China}
\author{Xian-Hui Zhong}
\email{zhongxh@hunnu.edu.cn}
\affiliation{Department of Physics, Hunan Normal University, and Key Laboratory of Low-Dimensional Quantum Structures and Quantum Control of Ministry of Education, Changsha 410081, China}
\affiliation{Synergetic Innovation Center for Quantum Effects and Applications (SICQEA), Hunan Normal University, Changsha 410081, China}

\begin{abstract}
Inspired by the newly observed $B_{s0}^*(5700)^0$, whose mass agrees well with our previous unquenched prediction for $B_s(1^3P_0)$ [Phys.~Rev.~D \textbf{109}, 116006 (2024)], we study the mass spectra and OZI-allowed two-body strong decays of the higher $3S$-, $2P$-, $2D$-, and $1F$-wave $B$ and $B_s$ mesons in the same unquenched quark model.
Our results show that coupled-channel effects significantly lower the masses and modify the relative level spacings.
The scalar state $B(2^3P_0)$ is strongly affected: its mass is lowered well below the quenched quark model predictions, and it mainly decays into $B\pi$.
Furthermore, the mass ordering of the vector states $B(2^3D_1)$ and $B(3^3S_1)$ is reversed, and the lighter $B(2^3D_1)$ state mainly decays into $B_1(5721)\pi$.
To test these results, the $B(2^3P_0)$ and $B(2^3D_1)$ states are worth searching for in the $B\pi$ and $B_1(5721)\pi$ channels, respectively.
In addition, some higher excitations are predicted to be relatively narrow and can be searched for in the $B^{(*)}\pi$, $B^{(*)}K$, $B_1(5721)\pi/K$, and $B_2^*(5747)\pi/K$ channels.
These predictions provide useful references for future experimental searches and help clarify the role of coupled-channel effects in the heavy-light meson spectrum.
\end{abstract}


\maketitle

\section{Introduction}

Recently, the LHCb Collaboration observed a new narrow bottom-strange state in the $B_s^0\pi^0$ invariant mass spectrum~\cite{LHCb:2026knl}.
Its mass and width were measured to be $M_{exp} = 5698.9\pm1.5\pm0.6~\mathrm{MeV}$ and $\Gamma_{exp} < 9.8~\mathrm{MeV}$, respectively.
Given its decay into two pseudoscalar mesons, the state is most naturally identified as the long-awaited lowest-lying $0^+$ scalar $B_{s}^0$ meson, denoted by $B_{s0}^*(5700)^0$.
It lies about $75~\mathrm{MeV}$ below the $BK$ threshold, whose mass is notably below the value $\sim5.8$ GeV predicted
in the conventional quark models.

By including coupled-channel effects in an unquenched quark model framework~\cite{Ni:2023lvx} for the $1^3P_0$ state, we have not only successfully explained the low mass nature of $D_{s0}^*(2317)^+$, but also predicted its bottom-strange counterpart with a mass of $M\simeq5693~\mathrm{MeV}$, which is in good agreement with the measured mass of $B_{s0}^*(5700)^0$.
In this unquenched picture, the strong $S$-wave coupling between the bare $b\bar s$ core and the $BK$ continuum induces a negative mass shift of about $80~\mathrm{MeV}$, which pulls the physical mass below the $BK$ threshold.

The coupled-channel effects on some low-lying bottom and bottom-strange mesons, such as the scalar state $B_s(1^3P_0)$ and axial-vector state $B_s(1P_1)$, have been investigated within various theoretical frameworks and approaches, including the chiral effective field theory and chiral unitary approach~\cite{Cheng:2017oqh,Kolomeitsev:2003ac,Guo:2006fu,Guo:2006rp,Du:2017zvv},
unquenched quark models~\cite{Ortega:2016pgg,Ni:2023lvx}, Hamiltonian effective field theory~\cite{Yang:2022vdb}, lattice QCD~\cite{Lang:2015hza}, and so on.
However, for the higher excitations of the $B$ and $B_s$ mesons, systematical theoretical studies have mostly relied on the quenched approximation.
Their mass spectra have been predicted within various quenched quark models~\cite{Zeng:1994vj,DiPierro:2001dwf,Ebert:2009ua,Sun:2014wea,Godfrey:2016nwn,Liu:2016efm,Lu:2016bbk,Kher:2017mky,Asghar:2018tha,li:2021hss} as well as QCD sum rules~\cite{Zhou:2014ytp,Zhou:2015ywa}, while their strong decays were investigated mainly within the $^3P_0$ model~\cite{Sun:2014wea,Godfrey:2016nwn,Lu:2016bbk,Ferretti:2015rsa}, the chiral quark model~\cite{Zhong:2008kd,Xiao:2014ura,li:2021hss}, and heavy quark effective theory~\cite{Eichten:1993ub,Falk:1995th,Casalbuoni:1996pg,Colangelo:2005gb,Colangelo:2012xi,Wang:2014cta,Campanella:2018xev}.
In these calculations, the feedback of the decay channels on the bare masses was not taken into account.
However, since the higher states lie above many open channel thresholds, their couplings to the decay channels may induce non-negligible coupled-channel effects~\cite{Isgur:1999cd}, which can shift the bare masses and modify the mass spectrum behavior.
Consequently, the quenched predictions for the mass spectra need to be reexamined before they are used to guide searches for the missing higher excitations.

Recently, our unquenched quark model study has shown that the coupled-channel effects are significant in the higher $3S$-, $2P$-, $2D$-, and $1F$-wave charmed and charmed-strange meson states~\cite{Ni:2026vjg}. 
Including the coupled-channel effects, the masses for most of these higher excitations are shifted down by tens to a few hundreds of MeV.
It should be mentioned that the newly observed $D_{s1}(2933)^+$ at LHCb~\cite{LHCb:2026sup}, which is much lower than conventional quark model predictions, however, can be assigned as the lower mixed axial-vector state $D_s(2P_1)$ within our unquenched quark model.
According to heavy quark flavor symmetry, the bottom and charmed mesons share similar chiral dynamics in the light quark sector.
Thus, comparable coupled-channel effects are also expected to occur in the higher $B$ and $B_s$ meson families.
However, a systematic study of the masses and strong decays of these higher states within the unquenched quark model framework is still lacking.
%

In this work, we extend our unquenched quark model framework~\cite{Ni:2023lvx,Ni:2026vjg} to systematically calculate the mass spectra and OZI-allowed two-body strong decays of the higher $3S$-, $2P$-, $2D$-, and $1F$-wave $B$ and $B_s$ mesons.
The mass shifts and strong decay widths are calculated consistently with the same transition amplitudes within the chiral quark model, and all parameters are adopted directly from our previous studies~\cite{Ni:2023lvx,Ni:2026vjg} without introducing additional free parameters.
In addition to the physical masses, we analyze the contributions of individual channels to the mass shifts and predict the total decay widths and dominant branching fractions.
These results provide useful references for searching for the higher $B$ and $B_s$ mesons in future experiments.

This paper is organized as follows.
In Sec.~\ref{sec:formalism}, we briefly describe the unquenched framework.
The numerical results and discussion are presented in Sec.~\ref{sec:results}, followed by a summary in Sec.~\ref{sec:summary}.

\section{Theoretical Framework}\label{sec:formalism}

In this work, the calculations are carried out within the unified unquenched quark model developed in Ref.~\cite{Ni:2023lvx}.

For a bare $Q\bar q$ meson in its rest frame, the Hamiltonian $\mathcal{H}_0$ includes the kinetic energies and an effective interaction $V(r)$,
\begin{equation}
\label{eq:bare-hamiltonian}
\mathcal{H}_{0}
=
\sqrt{\boldsymbol{p}_1^2+m_1^2}
+
\sqrt{\boldsymbol{p}_2^2+m_2^2}
+
V(r),
\end{equation}
where $m_1$ and $m_2$ are the constituent masses of the light and heavy quarks with momenta $\boldsymbol p_1$ and $\boldsymbol p_2$, respectively, and $r=|\boldsymbol r_1-\boldsymbol r_2|$ is the interquark distance.
The effective interaction $V(r)$ consists of the standard Cornell central potential and the spin-dependent interactions, which include the smeared contact term, the tensor force, and the spin-orbit force.
Their explicit forms and potential parameters are taken from Ref.~\cite{Ni:2023lvx}.

For a heavy-light meson system, because the constituent masses of the light and heavy quarks are unequal, the antisymmetric part of the spin-orbit potential, $H_{\mathrm{anti}}$, does not vanish and can cause a configuration mixing between the spin-triplet $n\,{}^3L_J$ and spin-singlet $n\,{}^1L_J$ states defined in the $L$-$S$ coupling scheme.
Thus, the mixed states $nL_J$ and $nL'_J$ are expressed as
\begin{equation}
\label{eq:state-mixing}
\begin{pmatrix}
nL_J\\
nL'_J
\end{pmatrix}
=
\begin{pmatrix}
\cos\theta_{nL} & \sin\theta_{nL}\\
-\sin\theta_{nL} & \cos\theta_{nL}
\end{pmatrix}
\begin{pmatrix}
n\,{}^1L_J\\
n\,{}^3L_J
\end{pmatrix},
\end{equation}
where $J=L=1,2,3,\cdots$, and $\theta_{nL}$ is the mixing angle.
In this work, $nL_J$ and $nL'_J$ correspond to the lower- and higher-mass mixed states, respectively, as often adopted in the literature.
The mixing angle $\theta_{nL}$ is perturbatively determined with the nondiagonal matrix element $\langle n\,{}^1L_J|H_{\mathrm{anti}}|n\,{}^3L_J\rangle$~\cite{Ni:2021pce,Ni:2023lvx}.

The bare mass $M_A$ and wave function $|A\rangle$ for each $Q\bar q$ state are obtained by solving the Schr\"{o}dinger equation with the bare Hamiltonian in Eq.~(\ref{eq:bare-hamiltonian}) via the Gaussian expansion method~\cite{Hiyama:2003cu}, with details given in Refs.~\cite{Ni:2021pce,Ni:2023lvx}.
By coupling this bare valence core $|A\rangle$ to the $BC$ continuum components, the physical state $|\Psi(M)\rangle$ is expanded as~\cite{Kalashnikova:2005ui,Barnes:2007xu,Lu:2016mbb,Ni:2023lvx}
\begin{equation}
\label{eq:physical-state-expansion}
|\Psi(M)\rangle
=
c_A|A\rangle
+
\sum_{BC}
\int
c_{BC}(\boldsymbol q)
|BC,\boldsymbol q\rangle
\,\mathrm{d}^3\boldsymbol q.
\end{equation}
Here, $c_A$ and $c_{BC}(\boldsymbol q)$ are the amplitudes of the bare core and the two-meson continuum state $|BC,\boldsymbol q\rangle$, respectively.
In the coupled space, the total Hamiltonian is $\mathcal{H} = \mathcal{H}_{0} + \mathcal{H}_{c} + \mathcal{H}_{I}$, where $\mathcal H_0|A\rangle=M_A|A\rangle$, $\mathcal H_c|BC,\boldsymbol q\rangle=E_{BC}(q)|BC,\boldsymbol q\rangle$ with $E_{BC}(q)=\sqrt{M_B^2+q^2}+\sqrt{M_C^2+q^2}$, and $\mathcal H_I$ is the coupling interaction.

Substituting Eq.~(\ref{eq:physical-state-expansion}) into the Schr\"{o}dinger equation $\mathcal H|\Psi(M)\rangle=M|\Psi(M)\rangle$ and eliminating $c_{BC}(\boldsymbol q)$ yields the mass equation~\cite{Kalashnikova:2005ui,Barnes:2007xu,Lu:2016mbb,Ni:2023lvx}
\begin{equation}
	M=M_A+\Sigma(M),
\end{equation}
with
\begin{equation}
\label{eq:self-energy-complex}
\Sigma(M)
=
\sum_{BC}
\int_0^\infty
\frac{
\overline{\left|\mathcal{M}_{A\to BC}(\boldsymbol{q})\right|^2}
}{
M-E_{BC}(q)+i\epsilon
}
q^2\,\mathrm{d}q.
\end{equation}
The averaged transition matrix element in Eq.~(\ref{eq:self-energy-complex}) is defined by
$\overline{\left|\mathcal{M}_{A\to BC}(\boldsymbol{q})\right|^2}
\equiv
\int
\overline{
	\left|
	\langle BC,\boldsymbol{q}|\mathcal{H}_{I}|A\rangle
	\right|^2
}
\,\mathrm{d}\Omega_{\boldsymbol{q}}$, where the overline denotes the spin average over the initial state and the spin sum over the final states.

The infinitesimal term $i\epsilon$ ($\epsilon\to 0^+$) fixes the analytic continuation above the open-channel threshold.
The real part of $\Sigma(M)$ gives the mass shift $\Delta M(M)$ due to the coupled-channel effects,
\begin{equation}
\label{eq:self-energy}
\Delta M(M)
=
\mathcal{P}
\sum_{BC}
\int_0^\infty
\frac{
\overline{\left|\mathcal{M}_{A\to BC}(\boldsymbol{q})\right|^2}
}{
M-E_{BC}(q)
}
q^2\,\mathrm{d}q,
\end{equation}
where $\mathcal{P}$ denotes the Cauchy principal value, while its imaginary part yields the partial width $\Gamma(A\to BC)$ for an open channel,
\begin{equation}
\label{eq:decay-width}
\Gamma(A\to BC)
=
2\pi
\frac{|\boldsymbol{q}|E_BE_C}{M}
\overline{
\left|
\mathcal{M}_{A\to BC}(\boldsymbol{q})
\right|^2
},
\end{equation}
where $\boldsymbol q$ is the on-shell relative momentum of $BC$ in the rest frame of $A$.
Both open and closed two-meson channels contribute to the self-energy in Eq.~(\ref{eq:self-energy}), whereas only kinematically open channels contribute to the decay widths in Eq.~(\ref{eq:decay-width}).
To remove the large overall mass shift induced by distant continuum states, we adopt the once-subtracted self-energy following Refs.~\cite{Pennington:2007xr,Zhou:2011sp,Duan:2021alw,Ni:2023lvx,Deng:2023mza,Ni:2025gvx},
\begin{align}
\label{eq:subtracted-self-energy}
\Delta M(M,M_0)
&\equiv
\Delta M(M)-\Delta M(M_0) \nonumber \\
&=
\mathcal{P}
\sum_{BC}
\int_0^\infty
\frac{
(M_0-M)\overline{\left|\mathcal{M}_{A\to BC}(\boldsymbol{q})\right|^2}
}{
[M-E_{BC}(q)][M_0-E_{BC}(q)]
}
q^2\,\mathrm{d}q.
\end{align}
The physical mass $M_{\rm phy}$ is then obtained from $M_{\rm phy}=M_A+\Delta M(M_{\rm phy},M_0)$.
As in Ref.~\cite{Ni:2023lvx}, the subtraction point $M_0$ is chosen as the mass of the corresponding ground state in the $B$ or $B_s$ family.
Through this subtraction, the smooth contributions from the distant virtual channels are absorbed into the bare mass, while the threshold effects from the nearby open channels are preserved in the mass shift.

The transition matrix elements are calculated with $\mathcal H_I$ in the chiral quark model, which has been widely applied in hadron strong decays~\cite{Zhong:2008kd,Xiao:2014ura,Zhong:2007gp,Wang:2017kfr,Ni:2021pce,li:2021hss}.
For light pseudoscalar meson emission, the quark-meson interaction is taken as~\cite{Manohar:1983md,Li:1994cy,Li:1997gd,Zhao:2002id}
\begin{equation}
\label{eq:chiral-lagrangian}
\mathcal{L}_{P}
=
\sum_j
\frac{1}{f_m}
\bar{\psi}_j
\gamma_{\mu}
\gamma_{5}
\psi_j
\vec{\tau}\cdot\partial^{\mu}\vec{\phi}_m,
\end{equation}
where $\psi_j$ is the $j$-th light quark field, $\phi_m$ denotes the emitted pseudoscalar field, $\tau$ is the isospin operator, and $f_m$ is the corresponding decay constant.
To match the nonrelativistic wave functions, the effective transition operator is derived by performing a Foldy-Wouthuysen reduction on the chiral Lagrangian in Eq.~(\ref{eq:chiral-lagrangian})~\cite{Ni:2023lvx},
\begin{equation}
	\mathcal H_I=\mathcal H_I^{\mathrm{NR}}+\mathcal H_I^{\mathrm{RC}}.
\end{equation}
The leading nonrelativistic operator $\mathcal H_I^{\mathrm{NR}}$ is given by
\begin{equation}
	\label{eq:transition-operator-nr}
	\mathcal{H}_{I}^{\mathrm{NR}}
	=
	g
	\sum_j
	\left[
	\mathcal{G}\,
	\boldsymbol{\sigma}_{j}\cdot\boldsymbol{q}
	+
	\frac{\omega_m}{2\mu_q}
	\boldsymbol{\sigma}_{j}\cdot\boldsymbol{p}_{j}
	\right]
	I_j\varphi_m,
\end{equation}
while the relativistic correction $\mathcal H_I^{\mathrm{RC}}$ is given by
\begin{align}
	\label{eq:transition-operator-rc}
	\mathcal{H}_{I}^{\mathrm{RC}}
	&=
	-
	\frac{g}{32\mu_q^2}
	\sum_j
	\Big[
	m_{\mathbb{P}}^2
	\boldsymbol{\sigma}_{j}\cdot\boldsymbol{q}
	\nonumber \\
	&\qquad
	+
	2\boldsymbol{\sigma}_{j}\cdot
	\left(
	\boldsymbol{q}-2\boldsymbol{p}_{j}
	\right)
	\times
	\left(
	\boldsymbol{q}\times\boldsymbol{p}_{j}
	\right)
	\Big]
	I_j\varphi_m.
\end{align}
Here $\boldsymbol p_j$ and $\boldsymbol\sigma_j$ are the internal momentum and spin operators of the active light quark, while $\boldsymbol q$ and $\omega_m$ are the momentum and energy of the emitted pseudoscalar meson with mass $m_{\mathbb P}$.
The plane-wave factor is $\varphi_m=\exp(-i\boldsymbol q\cdot\boldsymbol r_j)$, and $I_j$ is the SU(3) flavor operator~\cite{Li:1997gd}.
The transition factors are $g=\frac{\delta}{f_m}\sqrt{(E_i+M_i)(E_f+M_f)}$ and $\mathcal{G}=-\left(1+\frac{\omega_m}{E_f+M_f}+\frac{\omega_m}{2m'_j}\right)$, where $M_{i,f}$ and $E_{i,f}$ are the masses and energies of the initial/final heavy-light mesons, and $\mu_q=(1/m_j+1/m'_j)^{-1}$ is the reduced mass of the initial/final active quarks.

For decay channels involving a light vector meson, the quark--vector meson interaction is given by~\cite{Zhao:1998fn,Zhao:2000tb,Ni:2021pce}
\begin{equation}
\label{eq:vector-lagrangian}
\mathcal{L}_{V}
=
\sum_j
\bar{\psi}_j
\left(
a\gamma_{\mu}
+
\frac{i b}{2m_j}
\sigma_{\mu\nu}q^\nu
\right)
V^\mu
\psi_j,
\end{equation}
where $V^\mu$ is the emitted vector meson field, and $a$ and $b$ are the vector and tensor coupling strengths, respectively. After the nonrelativistic reduction, the transverse and longitudinal transition operators are~\cite{Zhao:1998fn}
\begin{align}
\label{eq:vector-transition-transverse}
\mathcal{H}_{I}^{V,T}
&=
\sum_j
\left[
i\frac{b'}{2m_q}
\boldsymbol{\sigma}_j\cdot
\left(
\boldsymbol q\times\boldsymbol\epsilon
\right)
+
\frac{a}{2\mu_q}
\boldsymbol p_j\cdot\boldsymbol\epsilon
\right]
I_j\varphi_m,
\end{align}
\begin{align}
\mathcal{H}_{I}^{V,L}
&=
\sum_j
\frac{aM_V}{|\boldsymbol q|}
I_j\varphi_m.\label{eq:vector-transition-longitudinal}
\end{align}
Here $\boldsymbol\epsilon$ is the transverse polarization vector, $M_V$ is the vector meson mass, $m_q$ is the light constituent quark mass used in the vector vertex, and $b'\equiv b-a$.

Since the quark-meson coupling is an effective low-energy interaction, a form factor is introduced to suppress the unphysical high-momentum contributions in the coupled-channel integral.
Following Ref.~\cite{Ni:2023lvx}, we adopt a Gaussian form factor to regularize the transition matrix element,
\begin{equation}
\label{eq:gaussian-suppression}
\langle BC,\boldsymbol{q}|\mathcal{H}_{I}|A\rangle
\rightarrow
\langle BC,\boldsymbol{q}|
\mathcal{H}_{I}
e^{-\frac{q^2}{2\Lambda^2}}
|A\rangle.
\end{equation}
Similar form factors have been widely used in coupled-channel calculations~\cite{Silvestre-Brac:1991qqx,Ortega:2016pgg,Ortega:2016mms,Yang:2021tvc,Yang:2022vdb}.
The cutoff parameter is fixed at $\Lambda=0.78~\mathrm{GeV}$ around the chiral symmetry breaking scale $\Lambda_{\chi{\rm SB}}\sim 1~\mathrm{GeV}$~\cite{Manohar:1983md}, which was determined by the low-lying states in Ref.~\cite{Ni:2023lvx} and is retained here without readjustment.
Other transition parameters are also taken directly from Ref.~\cite{Ni:2023lvx}: $\delta=0.557$, $f_\pi=132~\mathrm{MeV}$, $f_K=f_\eta=160~\mathrm{MeV}$, $\mu_q=225~\mathrm{MeV}$, and $a=-1.7$, $b'=2.5$ for vector meson emissions~\cite{Zhao:1998rt,Zhao:1998fn}.
The masses of the final-state mesons are taken from the Review of Particle Physics (RPP)~\cite{ParticleDataGroup:2024cfk} when experimentally available, while the unobserved ones are taken from our quark model predictions.
Because all parameters in the bare Hamiltonian and transition operators have been completely fixed by the low-lying states, the mass spectra and decay widths obtained below are genuine parameter-free predictions.

\begin{table*}[t]
	\caption{Mass spectra of the higher excited bottom and bottom-strange mesons in quenched and unquenched quark models.
		The quenched results labeled Ours correspond to $M_{\rm Q}$.
		For our unquenched results, $M_A$, $\Delta M$, and $M_{\rm phy}$ denote the bare mass, the diagonal mass shift induced by coupled-channel effects, and the corresponding physical mass, respectively.
		Mixed states with $J=L$ are ordered by increasing mass within each model.
		Ellipses denote unreported results. All masses are in MeV.}
	\label{tab:higher_bbs_ccef_spectrum}
	\setlength{\tabcolsep}{2.0pt}
	\renewcommand{\arraystretch}{1.05}
	\begin{tabular*}{\textwidth}{@{\extracolsep{\fill}}lcccccccccc@{}}
		\hline\hline
		\multirow{2}{*}{State} & \multirow{2}{*}{$J^P$} & \multicolumn{6}{c}{Quenched} & \multicolumn{3}{c}{Unquenched} \\
		\cmidrule(lr){3-8}\cmidrule(lr){9-11}
		& & EFG~\cite{Ebert:2009ua} & LPW~\cite{Lu:2016bbk} & GI~\cite{Godfrey:2016nwn} & LY~\cite{Liu:2015lka} & AMS~\cite{Asghar:2018tha} & Ours & $M_A$ & $\Delta M$ & $M_{\rm phy}$ \\
		\hline
		\multicolumn{11}{@{}l}{\textit{$B$ mesons}} \\
		$B(3^1S_0)$ & $0^-$ & 6379 & 6369 & 6335 & ... & 6288 & 6303 & 6315 & -209 & 6106 \\
		$B(3^3S_1)$ & $1^-$ & 6387 & 6391 & 6355 & ... & 6308 & 6320 & 6333 & -126 & 6207 \\
		$B(2^3P_0)$ & $0^+$ & 6221 & 6145 & 6213 & 6070 & 6129 & 6145 & 6156 & -271 & 5885 \\
		$B(2P_1)$ & $1^+$ & 6209 & 6185 & 6197 & 6137 & 6161 & 6178 & 6192 & -246 & 5946 \\
		$B(2P_1^\prime)$ & $1^+$ & 6281 & 6241 & 6228 & 6196 & 6175 & 6192 & 6204 & -173 & 6030 \\
		$B(2^3P_2)$ & $2^+$ & 6260 & 6253 & 6213 & 6220 & 6190 & 6203 & 6218 & -114 & 6104 \\
		$B(2^3D_1)$ & $1^-$ & 6534 & 6497 & 6475 & 6422 & 6384 & 6423 & 6438 & -284 & 6154 \\
		$B(2D_2)$ & $2^-$ & 6528 & 6435 & 6450 & 6417 & 6391 & 6408 & 6429 & -208 & 6221 \\
		$B(2D'_2)$ & $2^-$ & 6554 & 6513 & 6486 & 6446 & 6393 & 6461 & 6478 & -129 & 6349 \\
		$B(2^3D_3)$ & $3^-$ & 6542 & 6444 & 6460 & 6430 & 6395 & 6425 & 6446 & -103 & 6343 \\
		$B(1^3F_2)$ & $2^+$ & 6412 & 6383 & 6387 & 6374 & 6259 & 6377 & 6397 & -162 & 6235 \\
		$B(1F_3)$ & $3^+$ & 6391 & 6236 & 6358 & 6296 & 6249 & 6272 & 6296 & -134 & 6162 \\
		$B(1F'_3)$ & $3^+$ & 6420 & 6393 & 6396 & 6383 & 6264 & 6395 & 6415 & -106 & 6309 \\
		$B(1^3F_4)$ & $4^+$ & 6380 & 6243 & 6364 & 6302 & 6252 & 6282 & 6307 & -82 & 6225 \\
		\hline
		\multicolumn{11}{@{}l}{\textit{$B_s$ mesons}} \\
		$B_s(3^1S_0)$ & $0^-$ & 6467 & 6415 & 6410 & ... & 6305 & 6386 & 6394 & -108 & 6285 \\
		$B_s(3^3S_1)$ & $1^-$ & 6475 & 6435 & 6429 & ... & 6319 & 6404 & 6412 & -82 & 6329 \\
		$B_s(2^3P_0)$ & $0^+$ & 6318 & 6203 & 6279 & 6109 & 6160 & 6224 & 6227 & -119 & 6109 \\
		$B_s(2P_1)$ & $1^+$ & 6321 & 6241 & 6279 & 6187 & 6186 & 6264 & 6271 & -67 & 6204 \\
		$B_s(2P_1^\prime)$ & $1^+$ & 6345 & 6297 & 6296 & 6289 & 6196 & 6271 & 6276 & -68 & 6208 \\
		$B_s(2^3P_2)$ & $2^+$ & 6359 & 6309 & 6295 & 6308 & 6208 & 6289 & 6297 & -55 & 6242 \\
		$B_s(2^3D_1)$ & $1^-$ & 6629 & 6527 & 6542 & 6460 & 6390 & 6498 & 6501 & -108 & 6393 \\
		$B_s(2D_2)$ & $2^-$ & 6625 & 6492 & 6526 & 6483 & 6395 & 6493 & 6502 & -66 & 6436 \\
		$B_s(2D'_2)$ & $2^-$ & 6651 & 6542 & 6553 & 6499 & 6397 & 6535 & 6540 & -70 & 6470 \\
		$B_s(2^3D_3)$ & $3^-$ & 6637 & 6500 & 6535 & 6512 & 6399 & 6509 & 6519 & -42 & 6477 \\
		$B_s(1^3F_2)$ & $2^+$ & 6501 & 6412 & 6454 & 6405 & 6273 & 6444 & 6449 & -41 & 6408 \\
		$B_s(1F_3)$ & $3^+$ & 6468 & 6313 & 6425 & 6374 & 6265 & 6358 & 6368 & -30 & 6337 \\
		$B_s(1F'_3)$ & $3^+$ & 6515 & 6422 & 6462 & 6416 & 6277 & 6461 & 6468 & -60 & 6407 \\
		$B_s(1^3F_4)$ & $4^+$ & 6475 & 6319 & 6432 & 6380 & 6267 & 6368 & 6378 & -45 & 6333 \\
		\hline\hline
	\end{tabular*}
\end{table*}

\begin{figure*}[t]
	\centering
	\includegraphics[width=\textwidth]{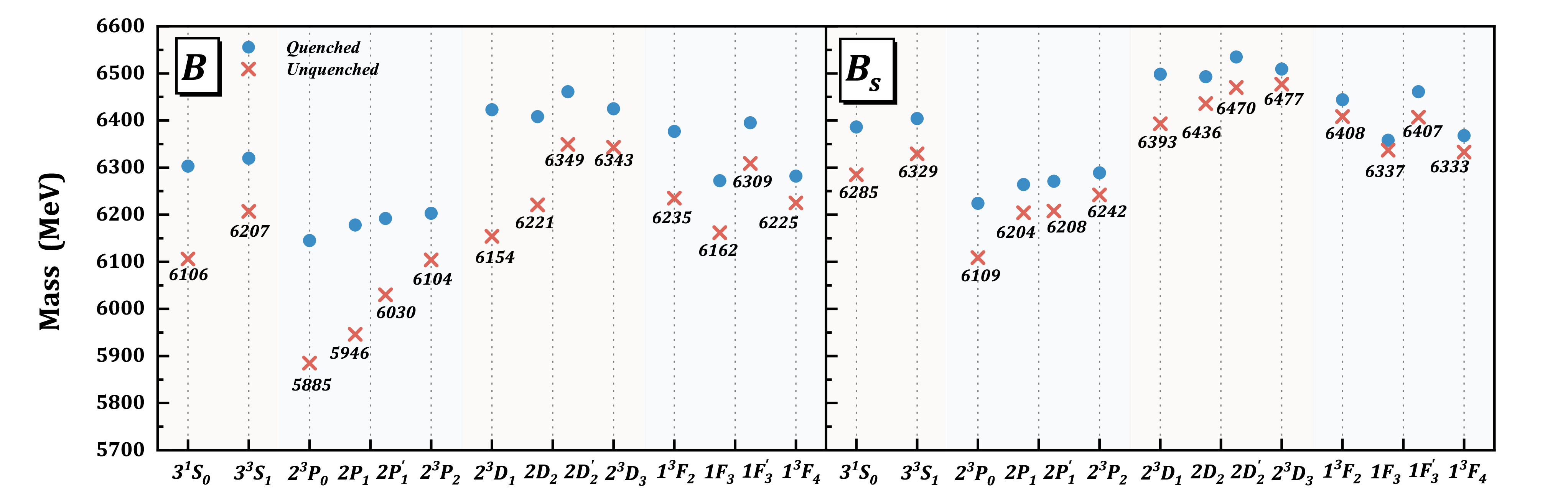}
	\caption{
		Mass spectra of the higher excited $B$ and $B_s$ mesons in the quenched and unquenched quark models.
		The blue circles and red crosses denote the quenched masses and the unquenched physical masses listed in Table~\ref{tab:higher_bbs_ccef_spectrum}, respectively.
	}
	\label{fig:higher_bbs_mass_spectrum}
\end{figure*}

\begin{figure*}[t]
	\centering
	\includegraphics[width=\textwidth]{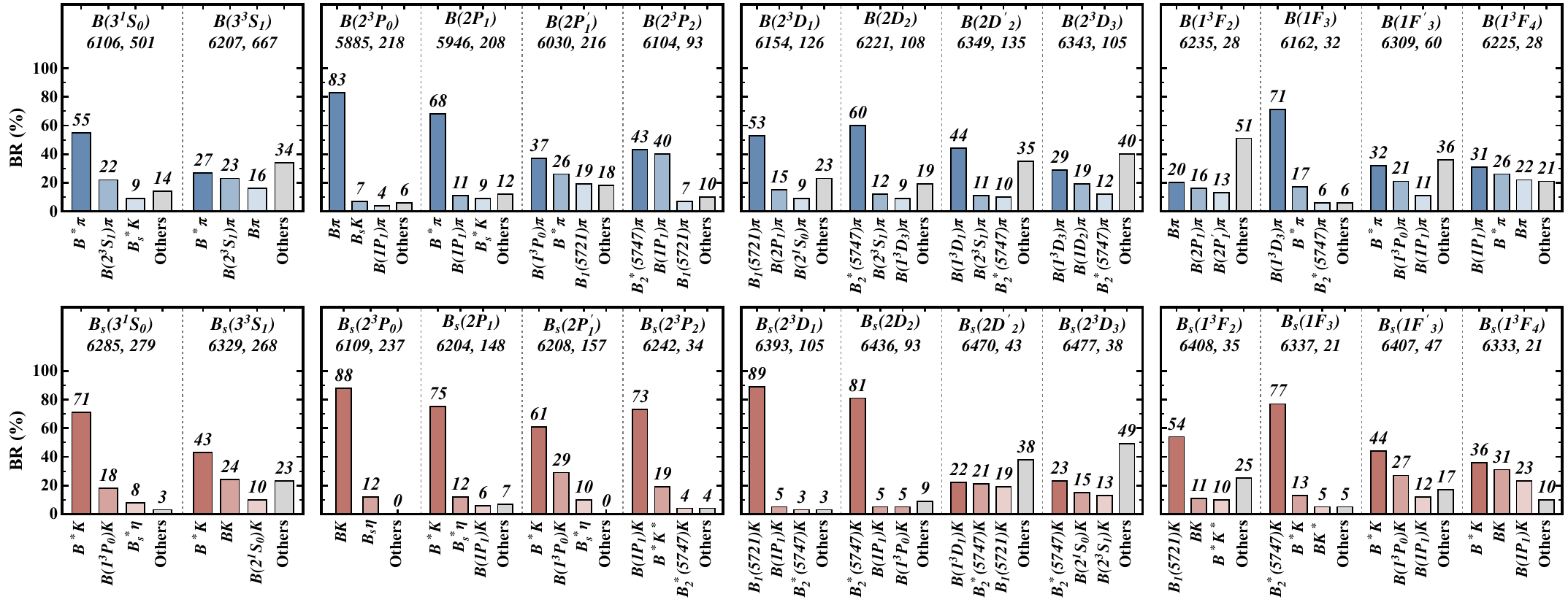}
	\caption{Predicted dominant OZI-allowed decay branching fractions of the higher $B$ (upper row) and $B_s$ (lower row) mesons.
		The two numbers below each state label are its physical mass and total strong decay width, respectively, in MeV.}
	\label{fig:higher_bbs_dominant_branching}
\end{figure*}

\section{Results and discussion}\label{sec:results}

The mass spectra of the higher $3S$-, $2P$-, $2D$-, and $1F$-wave $B$ and $B_s$ mesons calculated in both the quenched and unquenched quark models are listed in Table~\ref{tab:higher_bbs_ccef_spectrum}, together with the predictions of other quenched quark models~\cite{Ebert:2009ua,Lu:2016bbk,Godfrey:2016nwn,Liu:2015lka,Asghar:2018tha}, and are also plotted in Fig.~\ref{fig:higher_bbs_mass_spectrum}.
The physical masses, total strong decay widths, and dominant decay branching fractions are shown in Fig.~\ref{fig:higher_bbs_dominant_branching}, while the channel contributions to the mass shifts and partial widths are listed in Tables~\ref{tab:B_3S_channel_ccef}--\ref{tab:Bs_1F_channel_ccef} of the Appendix.

Compared with the quenched quark model predictions, coupled-channel effects lower the masses of all the higher $B$ and $B_s$ mesons considered here.
The mass shifts scatter about in the range of $(-300, -80)$ MeV for the $B$ mesons, while $(-120, -30)$ MeV for the $B_s$ mesons.
For the $B$ mesons, owing to the small pion mass, many pion-emission channels, such as $B^{(*)}\pi$, $B(2S)\pi$, $B(1P)\pi$, $B(1D)\pi$, and $B(2P)\pi$, are densely distributed in the mass region of the higher excitations.
Consequently, several of these channels lie close to the bare masses and couple strongly to the excited $B$ mesons, leading to sizable mass shifts.
By contrast, single-pion emission is forbidden by isospin conservation for the $B_s$ mesons, which can only couple to channels with kaon or $\eta$ emission.
Because the kaon and $\eta$ are much heavier than the pion and have larger decay constants ($f_{K,\eta}>f_\pi$), the allowed $K$- and $\eta$-emission channels have higher and more sparsely distributed thresholds, which usually lie farther from the bare masses, as well as weaker effective couplings than the $\pi$-emission channels.
Thus, the coupled-channel mass shifts are much more significant in the $B$ family than in the $B_s$ family.
A similar phenomenon was also found in the charmed sector, where the mass shifts are about $(-200, -50)$ MeV for the $D$ mesons and $(-90,-20)$ MeV for the $D_s$ mesons~\cite{Ni:2026vjg}.

Furthermore, the mass shifts depend strongly on the partial waves of the coupled channels and on the spacings between
the states and corresponding mass thresholds of coupled channels.
States coupling to nearby $S$-wave channels acquire large mass shifts, while states coupling to higher partial waves, such as the $2^3P_2$ and $1^3F_4$ states, receive much smaller shifts due to the centrifugal barrier.
Among the multiplets of the $B$ family, the $2P$ states are most strongly affected, with an average mass shift of about $-200~\mathrm{MeV}$.
Their mass shifts are $-271$, $-246$, $-173$, and $-114~\mathrm{MeV}$ for $B(2^3P_0)$, $B(2P_1)$, $B(2P_1^\prime)$, and $B(2^3P_2)$, respectively.
Because of these unequal shifts, the mass splitting of $B(2^3P_2)-B(2^3P_0)$, $\sim220~\mathrm{MeV}$, predicted in the
unquenched framework is significantly different from the value $\sim62~\mathrm{MeV}$ predicted within the quenched framework.
As a result, the three low-lying $B(2P)$ states are predicted at $5.89$--$6.03~\mathrm{GeV}$, which is
about $150$--$340~\mathrm{MeV}$ below the quenched quark model predictions as listed in Table~\ref{tab:higher_bbs_ccef_spectrum}.

For the scalar state $B(2^3P_0)$, the bare mass $M_A\simeq 6156~\mathrm{MeV}$ is comparable with the quenched quark
model predictions~\cite{Ebert:2009ua,Lu:2016bbk,Godfrey:2016nwn,Liu:2015lka,Asghar:2018tha}.
However, after including coupled-channel effects, its mass is significantly shifted down to the physical point $M_{\rm phy}\simeq 5885~\mathrm{MeV}$.
From Table~\ref{tab:B_2P_channel_ccef}, it is found that this large mass shift mainly arises from the $S$-wave couplings to the $B\pi$, $B(2^1S_0)\pi$, and $B_sK$ channels, which contribute $-103$, $-81$, and $-36~\mathrm{MeV}$, respectively.
Correspondingly, $B(2^3P_0)$ is predicted to be a broad state with $\Gamma\simeq 218~\mathrm{MeV}$, decaying predominantly into $B\pi$ with a branching fraction of $\sim 83\%$, as shown in Fig.~\ref{fig:higher_bbs_dominant_branching}, while its decay into $B^*\pi$ is forbidden by spin-parity conservation.
Therefore, $B(2^3P_0)$ provides a good opportunity to test the unquenched framework including coupled-channel effects.
If a broad $J^P=0^+$ resonance is to be observed around $5.9~\mathrm{GeV}$ with a dominant $B\pi$ decay and no $B^*\pi$ signal,
it would favor the unquenched predictions, whereas finding a scalar state around $6.1-6.2~\mathrm{GeV}$ would favor the quenched picture.
A partial-wave analysis of the $B\pi$ invariant mass spectrum around $5.9~\mathrm{GeV}$ is expected to be
carried out in future LHCb experiments.

While in the $B_s$ family, there are smaller mass shifts of around $(-120,-55)$~MeV for the bare $2P$ states.
The physical masses of the two mixed axial-vector states $B_s(2P_1)$ and $B_s(2P_1^\prime)$, 
$M_{\rm phy}\simeq 6204$ and $6208~\mathrm{MeV}$, are highly degenerate.
Fig.~\ref{fig:higher_bbs_dominant_branching} shows that both of these $1^+$ states have broad widths of $\Gamma\sim 150~\mathrm{MeV}$ and decay dominantly into $B^*K$, with branching fractions of $\sim 75\%$ and $\sim 61\%$, respectively.
Since their widths are much larger than their mass difference, they may appear as a single broad enhancement 
around $6.2~\mathrm{GeV}$ in the $B^*K$ invariant mass spectrum, or in the $BK$ spectrum through 
feed down.

Another notable consequence of coupled-channel effects is the mass order reversal of the two vector states $B(2^3D_1)$ and $B(3^3S_1)$, while it does not happen between the $B_s(2^3D_1)$ and $B_s(3^3S_1)$ states.
In the quenched quark models, $B(3^3S_1)$ lies about $100~\mathrm{MeV}$ below $B(2^3D_1)$, as can be seen 
from Table~\ref{tab:higher_bbs_ccef_spectrum} and Fig.~\ref{fig:higher_bbs_mass_spectrum}.
Including the coupled-channel effects, the mass shift of $B(2^3D_1)$, $-284~\mathrm{MeV}$, is much more 
significant than that of $B(3^3S_1)$, $-126~\mathrm{MeV}$, so that $B(2^3D_1)$ ($M_{\rm phy}\simeq 6154~\mathrm{MeV}$) 
lies below $B(3^3S_1)$ ($M_{\rm phy}\simeq 6207~\mathrm{MeV}$).
However, for the $B_s$ case, the physical $B_{s}(3^3S_1)$ ($M_{\rm phy}\simeq 6329~\mathrm{MeV}$) is still lower than $B_{s}(2^3D_1)$ ($M_{\rm phy}\simeq 6393~\mathrm{MeV}$) since the mass shifts are significantly reduced.
The details of the coupled-channel effects can be seen in Tables~\ref{tab:B_3S_channel_ccef} and~\ref{tab:B_2D_channel_ccef}.
It may be interesting to search for the vector state $B(2^3D_1)$ with mass of 
$M_{\rm phy}\simeq 6154~\mathrm{MeV}$ in future experiments, which provides a strong test for the unquenched quark model predictions. 
%
As illustrated in Fig.~\ref{fig:higher_bbs_dominant_branching}, this state has a moderate width of 
$\Gamma\simeq 126~\mathrm{MeV}$ and decays dominantly into $B_1(5721)\pi$ with a branching fraction of $\sim 53\%$.
Since $B_1(5721)$ is narrow and well established, the $B_1(5721)\pi$ provides a clean channel for future experimental observations.
By contrast, the other vector state $B(3^3S_1)$ is very broad ($\Gamma\simeq 667~\mathrm{MeV}$), thus, it may be difficult to identify in experiments.

Finally, several missing higher states have relatively narrow widths, as shown in Fig.~\ref{fig:higher_bbs_dominant_branching}, and may be observed in their dominant decay channels in future experiments.
For example, in the $B_s$ family, the $3^-$ state $B_s(2^3D_3)$ is predicted to have a mass of $M_{\rm phy}\simeq 6477~\mathrm{MeV}$ 
and a narrow width of $\Gamma\simeq 38~\mathrm{MeV}$, it may be observed in the $B_2^*(5747)K$ and $BK$ channels. 
Their branching fractions can reach up to $\sim 23\%$ and $\sim 13\%$, respectively.

\section{Summary}\label{sec:summary}

In this work, the mass spectra and OZI-allowed two-body strong decays of the higher $3S$-, $2P$-, $2D$-, and $1F$-wave $B$ and $B_s$ mesons are systematically investigated within an unquenched quark model.
It is found that the coupled-channel effects cause sizable state-dependent mass shifts, lowering the masses of the higher $B$ and $B_s$ mesons by $\sim80-300~\mathrm{MeV}$ and $\sim30-120~\mathrm{MeV}$, respectively, and significantly modifying the relative level spacings of the higher excitations.

The mass of the scalar state $B(2^3P_0)$ is lowered well below the quenched quark model predictions due to strong coupled-channel effects.
Because $B(2^3P_0)$ decays predominantly into $B\pi$, searching for this scalar state in the $B\pi$ invariant mass spectrum around $5.9~\mathrm{GeV}$ will provide a key test of the coupled-channel effects.
Furthermore, the strong coupled-channel effects reverse the mass ordering of the vector states $B(2^3D_1)$ and $B(3^3S_1)$, lowering $B(2^3D_1)$ below $B(3^3S_1)$.
Since $B(2^3D_1)$ decays dominantly into $B_1(5721)\pi$, searching for this state in the $B_1(5721)\pi$ channel around $6.15~\mathrm{GeV}$ will be very helpful to test this level reversal.

In addition, several missing higher excitations are predicted to have relatively narrow widths.
Besides the conventional channels $B^{(*)}\pi$ and $B^{(*)}K$, the $B_1(5721)\pi/K$ and $B_2^*(5747)\pi/K$ modes provide promising final states to search for them.
These theoretical predictions may provide useful references for future experimental searches and help understand the importance of coupled-channel effects in heavy-light systems.

\begin{acknowledgments}
This work is supported by the National Natural Science Foundation of China (Grants No.12221005,
No.12235018, No.12175065), the Chinese Academy of Sciences under Grant No. YSBR-101, and the
National Key Research and Development Program of China under Contract No. 2025YFA1613900.
\end{acknowledgments}

\appendix

\section{Channel contributions to mass shifts and strong decay widths}
\label{app:channel_decomposition}

The channel contributions to the mass shifts and OZI-allowed strong decay widths of the higher excited $B$ and $B_s$
mesons are listed in Tables~\ref{tab:B_3S_channel_ccef},
\ref{tab:Bs_3S_channel_ccef}, \ref{tab:B_2P_channel_ccef}, \ref{tab:Bs_2P_channel_ccef},
\ref{tab:B_2D_channel_ccef}, \ref{tab:Bs_2D_channel_ccef}, \ref{tab:B_1F_channel_ccef}, and
\ref{tab:Bs_1F_channel_ccef}.

\begin{table}[t!]
\caption{Mass shifts $\Delta M_i$ and OZI-allowed partial widths $\Gamma_i$ for the $B(3S)$ states. The numbers in square brackets in the column headers denote the bare mass $M_A$ and the physical mass $M_{\rm phy}$, respectively. All values are in MeV.}
\label{tab:B_3S_channel_ccef}
\setlength{\tabcolsep}{2.5pt}
\renewcommand{\arraystretch}{1.05}
\begin{tabular}{lcccc}
\hline\hline
  & \multicolumn{2}{c}{$B(3^1S_0)$} & \multicolumn{2}{c}{$B(3^3S_1)$} \\
\cline{2-3}\cline{4-5}
Channel & $\Delta M_i [6315]$ & $\Gamma_i [6106]$ & $\Delta M_i [6333]$ & $\Gamma_i [6207]$ \\
\hline
$B \pi$ & ... & ... & 10.7 & 104.3 \\
$B_s K$ & ... & ... & -4.5 & 31.2 \\
$B \eta$ & ... & ... & -1.3 & 13.4 \\
$B \eta'$ & ... & ... & -1.0 & ... \\
$B^* \pi$ & -24.8 & 277.3 & 6.5 & 178.4 \\
$B_s^* K$ & -20.5 & 42.9 & -10.7 & 45.6 \\
$B^* \eta$ & -8.1 & 21.7 & -3.7 & 20.7 \\
$B^* \eta'$ & -1.9 & ... & -1.5 & ... \\
$B(2^1S_0) \pi$ & ... & ... & -18.7 & 92.9 \\
$B(2^1S_0) \eta$ & ... & ... & -1.2 & ... \\
$B(2^3S_1) \pi$ & -72.4 & 110.6 & -42.8 & 152.9 \\
$B(1^3P_0) \pi$ & -35.5 & 38.4 & ... & ... \\
$B_s(1^3P_0) K$ & -11.8 & ... & ... & ... \\
$B(1^3P_0) \eta$ & -4.7 & ... & ... & ... \\
$B(1P_1) \pi$ & ... & ... & -15.9 & 12.8 \\
$B_s(1P_1) K$ & ... & ... & -5.0 & ... \\
$B(1P_1) \eta$ & ... & ... & -1.9 & 0.1 \\
$B_1(5721) \pi$ & ... & ... & -7.8 & 5.1 \\
$B_{s1}(5830) K$ & ... & ... & -2.2 & ... \\
$B_1(5721) \eta$ & ... & ... & -1.0 & ... \\
$B_2^*(5747) \pi$ & -13.3 & 9.4 & -7.0 & 8.7 \\
$B_2^*(5747) \eta$ & -1.3 & ... & -0.8 & ... \\
$B(1^3D_1) \pi$ & -2.1 & 0.3 & -0.6 & 1.0 \\
$B(1D_2) \pi$ & ... & ... & -2.0 & $2.3 \times 10^{-2}$ \\
$B(1D'_2) \pi$ & ... & ... & -5.7 & ... \\
$B(1^3D_3) \pi$ & -12.5 & ... & -6.7 & $7.3 \times 10^{-3}$ \\
$B \rho$ & $-3.2 \times 10^{-2}$ & $3.0 \times 10^{-2}$ & -0.3 & $4.2 \times 10^{-2}$ \\
$B \omega$ & $-1.2 \times 10^{-2}$ & $9.0 \times 10^{-3}$ & -0.1 & $1.1 \times 10^{-2}$ \\
$B_s K^*$ & $-2.7 \times 10^{-2}$ & ... & -0.1 & ... \\
$B^* \rho$ & -0.3 & $3.1 \times 10^{-3}$ & -0.4 & $3.2 \times 10^{-2}$ \\
$B^* \omega$ & -0.1 & ... & -0.1 & $1.3 \times 10^{-2}$ \\
$B_s^* K^*$ & -0.1 & ... & -0.2 & ... \\
\hline
$\mathrm{Total}$ & -209.4 & 500.6 & -125.9 & 667.1 \\
\hline\hline
\end{tabular}
\end{table}

\begin{table}[t!]
\caption{Mass shifts $\Delta M_i$ and OZI-allowed partial widths $\Gamma_i$ for the $B_s(3S)$ states. Notation and units are the same as in Table~\ref{tab:B_3S_channel_ccef}.}
\label{tab:Bs_3S_channel_ccef}
\setlength{\tabcolsep}{2.5pt}
\renewcommand{\arraystretch}{1.05}
\begin{tabular}{lcccc}
\hline\hline
  & \multicolumn{2}{c}{$B_s(3^1S_0)$} & \multicolumn{2}{c}{$B_s(3^3S_1)$} \\
\cline{2-3}\cline{4-5}
Channel & $\Delta M_i [6394]$ & $\Gamma_i [6285]$ & $\Delta M_i [6412]$ & $\Gamma_i [6329]$ \\
\hline
$B K$ & ... & ... & 4.4 & 64.1 \\
$B_s \eta$ & ... & ... & -0.5 & 9.3 \\
$B_s \eta'$ & ... & ... & -1.8 & $4.7 \times 10^{-2}$ \\
$B^* K$ & -16.2 & 198.3 & -0.3 & 115.2 \\
$B_s^* \eta$ & -5.1 & 23.6 & -2.1 & 15.3 \\
$B_s^* \eta'$ & -4.2 & ... & -2.7 & ... \\
$B(2^1S_0) K$ & ... & ... & -17.8 & 27.1 \\
$B(2^3S_1) K$ & -37.2 & ... & -31.9 & 14.0 \\
$B(1^3P_0) K$ & -25.0 & 49.1 & ... & ... \\
$B_s(1^3P_0) \eta$ & -5.3 & 4.7 & ... & ... \\
$B(1P_1) K$ & ... & ... & -10.8 & 13.2 \\
$B_s(1P_1) \eta$ & ... & ... & -1.6 & 0.3 \\
$B_1(5721) K$ & ... & ... & -7.3 & 3.8 \\
$B_{s1}(5830) \eta$ & ... & ... & -0.8 & ... \\
$B_2^*(5747) K$ & -13.6 & 3.1 & -7.3 & 4.9 \\
$B_{s2}^*(5840) \eta$ & -1.3 & ... & -0.7 & ... \\
$B K^*$ & $-3.2 \times 10^{-2}$ & 0.1 & -0.3 & $3.0 \times 10^{-2}$ \\
$B_s \phi$ & $-2.3 \times 10^{-2}$ & ... & $-7.7 \times 10^{-2}$ & ... \\
$B^* K^*$ & -0.4 & 0.1 & -0.4 & 0.1 \\
\hline
$\mathrm{Total}$ & -108.4 & 279.0 & -82.2 & 267.5 \\
\hline\hline
\end{tabular}
\end{table}

\begin{table*}[t!]
\caption{Mass shifts $\Delta M_i$ and OZI-allowed partial widths $\Gamma_i$ for the $B(2P)$ states. Notation and units are the same as in Table~\ref{tab:B_3S_channel_ccef}.}
\label{tab:B_2P_channel_ccef}
\setlength{\tabcolsep}{0.85pt}
\renewcommand{\arraystretch}{1.05}
\begin{tabular*}{\textwidth}{@{\extracolsep{\fill}}lcccccccc@{}}
\hline\hline
  & \multicolumn{2}{c}{$B(2^3P_0)$} & \multicolumn{2}{c}{$B(2P_1)$} & \multicolumn{2}{c}{$B(2P_1^\prime)$} & \multicolumn{2}{c}{$B(2^3P_2)$} \\
\cline{2-3}\cline{4-5}\cline{6-7}\cline{8-9}
Channel & $\Delta M_i [6156]$ & $\Gamma_i [5885]$ & $\Delta M_i [6192]$ & $\Gamma_i [5946]$ & $\Delta M_i [6204]$ & $\Gamma_i [6030]$ & $\Delta M_i [6218]$ & $\Gamma_i [6104]$ \\
\hline
$B \pi$ & -103.0 & 181.1 & ... & ... & ... & ... & -5.4 & 0.1 \\
$B_s K$ & -36.3 & 15.5 & ... & ... & ... & ... & -1.6 & $2.8 \times 10^{-2}$ \\
$B \eta$ & -14.1 & 9.1 & ... & ... & ... & ... & -0.7 & $4.7 \times 10^{-2}$ \\
$B^* \pi$ & ... & ... & -62.7 & 141.5 & -17.6 & 56.3 & -5.9 & 0.2 \\
$B_s^* K$ & ... & ... & -27.7 & 18.8 & -11.2 & 13.3 & -2.0 & 0.2 \\
$B^* \eta$ & ... & ... & -10.8 & 10.5 & -4.3 & 6.4 & -0.8 & 0.2 \\
$B(2^1S_0) \pi$ & -81.2 & ... & ... & ... & ... & ... & -5.8 & 2.5 \\
$B(2^3S_1) \pi$ & ... & ... & -77.2 & ... & -37.7 & 17.0 & -7.7 & 1.9 \\
$B(1^3P_0) \pi$ & ... & ... & -2.4 & 6.2 & -18.5 & 80.5 & ... & ... \\
$B_s(1^3P_0) K$ & ... & ... & -0.6 & ... & -2.6 & ... & ... & ... \\
$B(1^3P_0) \eta$ & ... & ... & -0.3 & ... & -1.4 & ... & ... & ... \\
$B(1P_1) \pi$ & -8.7 & 9.7 & -9.4 & 22.3 & -3.7 & 0.1 & -17.5 & 37.0 \\
$B(1P_1) \eta$ & ... & ... & ... & ... & ... & ... & -1.5 & ... \\
$B_1(5721) \pi$ & -26.3 & 2.1 & -0.9 & 0.1 & -19.0 & 41.3 & -7.3 & 6.5 \\
$B_2^*(5747) \pi$ & ... & ... & -21.1 & 9.0 & -11.0 & 1.2 & -14.1 & 39.9 \\
$B(1^3D_1) \pi$ & ... & ... & -8.8 & ... & -30.4 & ... & -1.3 & $1.3 \times 10^{-3}$ \\
$B(1D_2) \pi$ & ... & ... & -0.6 & ... & -1.0 & ... & -26.6 & ... \\
$B(1^3D_3) \pi$ & ... & ... & -19.8 & ... & -11.9 & ... & -10.4 & ... \\
$B \rho$ & ... & ... & -1.9 & ... & -0.5 & ... & -0.4 & $1.1 \times 10^{-2}$ \\
$B \omega$ & ... & ... & -0.6 & ... & -0.1 & ... & -0.1 & $1.6 \times 10^{-3}$ \\
$B^* \rho$ & -1.2 & ... & -1.1 & ... & -1.8 & ... & -3.9 & 4.6 \\
$B^* \omega$ & -0.4 & ... & -0.3 & ... & -0.6 & ... & -1.1 & ... \\
\hline
$\mathrm{Total}$ & -271.1 & 217.5 & -246.2 & 208.4 & -173.3 & 216.0 & -114.3 & 93.2 \\
\hline\hline
\end{tabular*}
\end{table*}

\begin{table*}[t!]
\caption{Mass shifts $\Delta M_i$ and OZI-allowed partial widths $\Gamma_i$ for the $B_s(2P)$ states. Notation and units are the same as in Table~\ref{tab:B_3S_channel_ccef}.}
\label{tab:Bs_2P_channel_ccef}
\setlength{\tabcolsep}{0.85pt}
\renewcommand{\arraystretch}{1.05}
\begin{tabular*}{\textwidth}{@{\extracolsep{\fill}}lcccccccc@{}}
\hline\hline
  & \multicolumn{2}{c}{$B_s(2^3P_0)$} & \multicolumn{2}{c}{$B_s(2P_1)$} & \multicolumn{2}{c}{$B_s(2P_1^\prime)$} & \multicolumn{2}{c}{$B_s(2^3P_2)$} \\
\cline{2-3}\cline{4-5}\cline{6-7}\cline{8-9}
Channel & $\Delta M_i [6227]$ & $\Gamma_i [6109]$ & $\Delta M_i [6271]$ & $\Gamma_i [6204]$ & $\Delta M_i [6276]$ & $\Gamma_i [6208]$ & $\Delta M_i [6297]$ & $\Gamma_i [6242]$ \\
\hline
$B K$ & -75.7 & 209.8 & ... & ... & ... & ... & -5.3 & 0.6 \\
$B_s \eta$ & -15.1 & 27.4 & ... & ... & ... & ... & -0.7 & $1.4 \times 10^{-2}$ \\
$B^* K$ & ... & ... & -21.6 & 111.2 & -19.3 & 96.0 & -5.9 & $1.3 \times 10^{-2}$ \\
$B_s^* \eta$ & ... & ... & -6.7 & 18.2 & -5.8 & 15.3 & -0.8 & $3.5 \times 10^{-2}$ \\
$B(2^1S_0) K$ & ... & ... & ... & ... & ... & ... & -5.3 & ... \\
$B(1^3P_0) K$ & ... & ... & -3.9 & 0.7 & -12.0 & 45.2 & ... & ... \\
$B_s(1^3P_0) \eta$ & ... & ... & -0.4 & ... & -0.9 & ... & ... & ... \\
$B(1P_1) K$ & -4.0 & ... & -6.5 & 8.8 & -2.7 & 0.3 & -14.2 & 25.2 \\
$B_1(5721) K$ & -22.5 & ... & -0.9 & ... & -11.6 & ... & -6.7 & 0.5 \\
$B_2^*(5747) K$ & ... & ... & -21.5 & ... & -12.4 & ... & -12.4 & 1.4 \\
$B K^*$ & ... & ... & -4.4 & 8.6 & -0.4 & 0.4 & -0.5 & $2.2 \times 10^{-2}$ \\
$B^* K^*$ & -1.6 & ... & -1.4 & ... & -3.0 & ... & -3.6 & 6.6 \\
\hline
$\mathrm{Total}$ & -118.8 & 237.2 & -67.2 & 147.5 & -68.1 & 157.1 & -55.4 & 34.4 \\
\hline\hline
\end{tabular*}
\end{table*}

\begin{table*}[t!]
\caption{Mass shifts $\Delta M_i$ and OZI-allowed partial widths $\Gamma_i$ for the $B(2D)$ states. Notation and units are the same as in Table~\ref{tab:B_3S_channel_ccef}.}
\label{tab:B_2D_channel_ccef}
\setlength{\tabcolsep}{0.85pt}
\renewcommand{\arraystretch}{1.05}
\begin{tabular*}{\textwidth}{@{\extracolsep{\fill}}lcccccccc@{}}
\hline\hline
  & \multicolumn{2}{c}{$B(2^3D_1)$} & \multicolumn{2}{c}{$B(2D_2)$} & \multicolumn{2}{c}{$B(2D'_2)$} & \multicolumn{2}{c}{$B(2^3D_3)$} \\
\cline{2-3}\cline{4-5}\cline{6-7}\cline{8-9}
Channel & $\Delta M_i [6438]$ & $\Gamma_i [6154]$ & $\Delta M_i [6429]$ & $\Gamma_i [6221]$ & $\Delta M_i [6478]$ & $\Gamma_i [6349]$ & $\Delta M_i [6446]$ & $\Gamma_i [6343]$ \\
\hline
$B \pi$ & -7.3 & 3.2 & ... & ... & ... & ... & -4.7 & 3.8 \\
$B_s K$ & -1.7 & 1.0 & ... & ... & ... & ... & -1.2 & 0.3 \\
$B \eta$ & -0.8 & 0.3 & ... & ... & ... & ... & -0.6 & 0.2 \\
$B \eta'$ & -0.3 & ... & ... & ... & ... & ... & -0.3 & $2.6 \times 10^{-3}$ \\
$B^* \pi$ & -3.3 & 0.8 & -9.7 & 0.5 & -8.0 & 2.4 & -5.9 & 3.1 \\
$B_s^* K$ & -0.7 & 0.3 & -3.4 & $2.6 \times 10^{-2}$ & -2.1 & $4.6 \times 10^{-2}$ & -1.4 & 0.2 \\
$B^* \eta$ & -0.3 & 0.1 & -1.5 & $6.7 \times 10^{-4}$ & -0.9 & $2.3 \times 10^{-2}$ & -0.7 & 0.1 \\
$B^* \eta'$ & -0.1 & ... & -0.4 & ... & -0.5 & $1.0 \times 10^{-2}$ & -0.3 & $2.6 \times 10^{-3}$ \\
$B(2^1S_0) \pi$ & -2.6 & 11.8 & ... & ... & ... & ... & -3.5 & 6.6 \\
$B_s(2^1S_0) K$ & -0.8 & ... & ... & ... & ... & ... & -0.4 & ... \\
$B(2^1S_0) \eta$ & -0.4 & ... & ... & ... & ... & ... & -0.3 & $2.0 \times 10^{-4}$ \\
$B(2^3S_1) \pi$ & -1.5 & 3.6 & -4.9 & 12.9 & -10.2 & 14.3 & -5.1 & 7.2 \\
$B_s(2^3S_1) K$ & ... & ... & ... & ... & -1.4 & ... & ... & ... \\
$B(2^3S_1) \eta$ & -0.2 & ... & -0.7 & ... & -0.8 & ... & -0.4 & ... \\
$B(1^3P_0) \pi$ & ... & ... & -0.7 & 5.9 & 1.3 & 2.9 & ... & ... \\
$B_s(1^3P_0) K$ & ... & ... & -0.1 & $4.6 \times 10^{-3}$ & -0.2 & 0.6 & ... & ... \\
$B(1^3P_0) \eta$ & ... & ... & -0.1 & $3.2 \times 10^{-2}$ & -0.1 & 0.5 & ... & ... \\
$B(1P_1) \pi$ & -2.3 & 8.5 & -1.3 & 8.4 & $2.8 \times 10^{-2}$ & 0.7 & 0.8 & 3.0 \\
$B_s(1P_1) K$ & -0.3 & ... & -0.2 & ... & -0.2 & 0.1 & -0.4 & 0.3 \\
$B(1P_1) \eta$ & -0.2 & ... & -0.1 & $1.7 \times 10^{-3}$ & -0.1 & 0.2 & -0.2 & 0.6 \\
$B_1(5721) \pi$ & -82.5 & 66.7 & -1.0 & $1.2 \times 10^{-2}$ & -1.8 & 12.4 & -3.4 & 4.0 \\
$B_{s1}(5830) K$ & -14.8 & ... & -0.3 & ... & -1.1 & $3.2 \times 10^{-2}$ & -0.7 & $2.5 \times 10^{-3}$ \\
$B_1(5721) \eta$ & -5.8 & ... & -0.1 & ... & -0.6 & 0.3 & -0.4 & $3.4 \times 10^{-2}$ \\
$B_2^*(5747) \pi$ & -1.8 & 1.1 & -52.0 & 65.0 & -8.6 & 13.6 & -3.7 & 12.6 \\
$B_{s2}^*(5840) K$ & -0.4 & ... & -10.2 & ... & -1.7 & 0.1 & -1.3 & $2.4 \times 10^{-3}$ \\
$B_2^*(5747) \eta$ & -0.2 & ... & -4.1 & ... & -0.9 & 0.2 & -0.7 & 0.2 \\
$B(1^3D_1) \pi$ & -13.3 & 9.2 & -1.6 & 0.8 & -43.6 & 58.9 & -0.7 & 0.2 \\
$B(1D_2) \pi$ & -0.7 & ... & -6.1 & 1.5 & -2.6 & 0.2 & -31.0 & 19.8 \\
$B(1D'_2) \pi$ & -15.2 & ... & -1.1 & ... & -14.6 & 10.9 & -4.6 & $2.8 \times 10^{-2}$ \\
$B(1^3D_3) \pi$ & -2.2 & $7.6 \times 10^{-7}$ & -21.8 & 9.7 & -8.9 & 0.2 & -13.4 & 30.9 \\
$B(2^3P_0) \pi$ & ... & ... & -2.0 & 0.6 & -1.5 & 5.7 & ... & ... \\
$B(2^3P_0) \eta$ & ... & ... & ... & ... & -0.2 & ... & ... & ... \\
$B(2P_1) \pi$ & -35.9 & 19.0 & -0.1 & $2.8 \times 10^{-2}$ & -0.1 & 0.5 & -6.2 & 8.2 \\
$B(2P_1^\prime) \pi$ & -82.5 & ... & -2.4 & $1.3 \times 10^{-2}$ & -4.1 & 2.1 & -5.5 & 0.2 \\
$B(2^3P_2) \pi$ & -1.9 & ... & -77.6 & ... & -13.6 & 4.2 & -5.5 & 0.2 \\
$B \rho$ & $-4.0 \times 10^{-2}$ & $3.2 \times 10^{-5}$ & -0.8 & 1.3 & $8.9 \times 10^{-3}$ & 0.8 & -0.2 & 0.2 \\
$B \omega$ & $-1.3 \times 10^{-2}$ & $1.6 \times 10^{-5}$ & -0.3 & 0.4 & $-8.4 \times 10^{-4}$ & 0.3 & -0.1 & $4.3 \times 10^{-2}$ \\
$B_s K^*$ & $-1.2 \times 10^{-2}$ & ... & -0.3 & ... & -0.1 & 0.2 & $-4.1 \times 10^{-2}$ & $1.0 \times 10^{-3}$ \\
$B^* \rho$ & -0.9 & 0.1 & -0.9 & 0.4 & -0.6 & 1.4 & -0.4 & 2.2 \\
$B^* \omega$ & -0.3 & $2.9 \times 10^{-2}$ & -0.3 & 0.1 & -0.2 & 0.5 & -0.1 & 0.7 \\
$B_s^* K^*$ & -0.3 & ... & -0.2 & ... & -0.3 & 0.1 & -0.3 & 0.2 \\
$B(1^3P_0) \rho$ & -1.8 & ... & $-3.6 \times 10^{-2}$ & ... & -0.3 & $4.1 \times 10^{-5}$ & $-2.1 \times 10^{-2}$ & ... \\
$B(1^3P_0) \omega$ & -0.6 & ... & $-1.2 \times 10^{-2}$ & ... & -0.1 & ... & $-6.8 \times 10^{-3}$ & ... \\
$B(1P_1) \rho$ & -0.4 & ... & -1.7 & ... & -0.3 & ... & -0.4 & ... \\
$B(1P_1) \omega$ & ... & ... & ... & ... & -0.1 & ... & -0.1 & ... \\
\hline
$\mathrm{Total}$ & -284.3 & 125.6 & -208.0 & 107.6 & -129.1 & 134.6 & -103.4 & 104.9 \\
\hline\hline
\end{tabular*}
\end{table*}

\begin{table*}[t!]
\caption{Mass shifts $\Delta M_i$ and OZI-allowed partial widths $\Gamma_i$ for the $B_s(2D)$ states. Notation and units are the same as in Table~\ref{tab:B_3S_channel_ccef}.}
\label{tab:Bs_2D_channel_ccef}
\setlength{\tabcolsep}{0.85pt}
\renewcommand{\arraystretch}{1.05}
\begin{tabular*}{\textwidth}{@{\extracolsep{\fill}}lcccccccc@{}}
\hline\hline
  & \multicolumn{2}{c}{$B_s(2^3D_1)$} & \multicolumn{2}{c}{$B_s(2D_2)$} & \multicolumn{2}{c}{$B_s(2D'_2)$} & \multicolumn{2}{c}{$B_s(2^3D_3)$} \\
\cline{2-3}\cline{4-5}\cline{6-7}\cline{8-9}
Channel & $\Delta M_i [6501]$ & $\Gamma_i [6393]$ & $\Delta M_i [6502]$ & $\Gamma_i [6436]$ & $\Delta M_i [6540]$ & $\Gamma_i [6470]$ & $\Delta M_i [6519]$ & $\Gamma_i [6477]$ \\
\hline
$B K$ & -4.5 & 0.1 & ... & ... & ... & ... & -4.3 & 5.0 \\
$B_s \eta$ & -0.9 & $2.0 \times 10^{-3}$ & ... & ... & ... & ... & -0.5 & 0.4 \\
$B_s \eta'$ & -0.8 & $1.0 \times 10^{-2}$ & ... & ... & ... & ... & -0.5 & $1.0 \times 10^{-3}$ \\
$B^* K$ & -2.3 & $7.0 \times 10^{-3}$ & -3.5 & 0.2 & -7.1 & 3.0 & -5.5 & 4.3 \\
$B_s^* \eta$ & -0.4 & $5.0 \times 10^{-3}$ & -1.1 & 0.1 & -0.9 & 0.1 & -0.6 & 0.3 \\
$B_s^* \eta'$ & -0.3 & $5.0 \times 10^{-3}$ & -1.4 & 0.1 & -1.0 & $2.2 \times 10^{-2}$ & -0.7 & $3.0 \times 10^{-3}$ \\
$B(2^1S_0) K$ & -3.0 & 0.1 & ... & ... & ... & ... & -4.3 & 5.8 \\
$B_s(2^1S_0) \eta$ & -0.4 & ... & ... & ... & ... & ... & -0.3 & $5.0 \times 10^{-5}$ \\
$B(2^3S_1) K$ & -1.7 & $5.5 \times 10^{-2}$ & -6.0 & 0.1 & -10.9 & 7.9 & -5.7 & 5.1 \\
$B_s(2^3S_1) \eta$ & ... & ... & ... & ... & -0.7 & ... & -0.4 & ... \\
$B(1^3P_0) K$ & ... & ... & -0.6 & 4.2 & 0.7 & 0.9 & ... & ... \\
$B_s(1^3P_0) \eta$ & ... & ... & -0.1 & 0.1 & $-4.2 \times 10^{-2}$ & 0.2 & ... & ... \\
$B(1P_1) K$ & -1.7 & 5.7 & -1.0 & 4.8 & 0.1 & 0.9 & 0.6 & 2.0 \\
$B_s(1P_1) \eta$ & -0.2 & 0.1 & -0.1 & 0.1 & -0.1 & 0.2 & -0.1 & 0.4 \\
$B_1(5721) K$ & -71.8 & 93.4 & -0.9 & 0.1 & -2.7 & 8.1 & -3.5 & 3.0 \\
$B_{s1}(5830) \eta$ & -8.5 & 1.8 & -0.1 & $3.0 \times 10^{-3}$ & -0.5 & 0.2 & -0.3 & $4.4 \times 10^{-2}$ \\
$B_2^*(5747) K$ & -2.1 & 2.6 & -41.6 & 75.3 & -8.3 & 9.0 & -4.7 & 8.9 \\
$B_{s2}^*(5840) \eta$ & -0.2 & $1.4 \times 10^{-4}$ & -5.8 & 3.0 & -0.8 & 0.2 & -0.7 & 0.2 \\
$B(1^3D_1) K$ & -4.9 & ... & -1.4 & ... & -28.6 & 9.3 & -0.4 & $4.0 \times 10^{-3}$ \\
$B(1^3D_3) K$ & ... & ... & ... & ... & -8.3 & ... & -8.9 & ... \\
$B K^*$ & -0.1 & $2.0 \times 10^{-3}$ & -1.0 & 2.8 & $9.0 \times 10^{-3}$ & 0.8 & -0.3 & 0.2 \\
$B_s \phi$ & $-1.4 \times 10^{-2}$ & $9.0 \times 10^{-9}$ & -0.3 & 0.1 & -0.1 & 0.2 & $-4.1 \times 10^{-2}$ & $1.0 \times 10^{-3}$ \\
$B^* K^*$ & -1.3 & 0.8 & -1.3 & 1.4 & -0.7 & 1.4 & -0.4 & 2.3 \\
$B_s^* \phi$ & -0.3 & ... & -0.3 & $2.7 \times 10^{-4}$ & -0.2 & 0.1 & -0.2 & 0.2 \\
$B(1^3P_0) K^*$ & -2.8 & ... & $-4.3 \times 10^{-2}$ & ... & -0.4 & $1.4 \times 10^{-5}$ & $-3.2 \times 10^{-2}$ & $1.0 \times 10^{-3}$ \\
\hline
$\mathrm{Total}$ & -108.3 & 104.6 & -66.4 & 92.5 & -70.4 & 42.5 & -41.7 & 38.1 \\
\hline\hline
\end{tabular*}
\end{table*}

\begin{table*}[t!]
\caption{Mass shifts $\Delta M_i$ and OZI-allowed partial widths $\Gamma_i$ for the $B(1F)$ states. Notation and units are the same as in Table~\ref{tab:B_3S_channel_ccef}.}
\label{tab:B_1F_channel_ccef}
\setlength{\tabcolsep}{0.85pt}
\renewcommand{\arraystretch}{1.05}
\begin{tabular*}{\textwidth}{@{\extracolsep{\fill}}lcccccccc@{}}
\hline\hline
  & \multicolumn{2}{c}{$B(1^3F_2)$} & \multicolumn{2}{c}{$B(1F_3)$} & \multicolumn{2}{c}{$B(1F'_3)$} & \multicolumn{2}{c}{$B(1^3F_4)$} \\
\cline{2-3}\cline{4-5}\cline{6-7}\cline{8-9}
Channel & $\Delta M_i [6397]$ & $\Gamma_i [6235]$ & $\Delta M_i [6296]$ & $\Gamma_i [6162]$ & $\Delta M_i [6415]$ & $\Gamma_i [6309]$ & $\Delta M_i [6307]$ & $\Gamma_i [6225]$ \\
\hline
$B \pi$ & -0.4 & 5.7 & ... & ... & ... & ... & -3.5 & 6.1 \\
$B_s K$ & -0.2 & 0.4 & ... & ... & ... & ... & -0.8 & 0.4 \\
$B \eta$ & -0.1 & 0.3 & ... & ... & ... & ... & -0.5 & 0.4 \\
$B \eta'$ & $-3.8 \times 10^{-2}$ & ... & ... & ... & ... & ... & -0.2 & ... \\
$B^* \pi$ & -0.4 & 3.1 & -1.1 & 5.3 & -9.5 & 19.3 & -5.1 & 7.3 \\
$B_s^* K$ & -0.2 & 0.1 & -0.3 & 0.2 & -2.5 & 1.7 & -1.1 & 0.4 \\
$B^* \eta$ & -0.1 & 0.1 & -0.2 & 0.1 & -1.4 & 1.5 & -0.6 & 0.4 \\
$B^* \eta'$ & $-2.7 \times 10^{-2}$ & ... & -0.1 & ... & -0.6 & $1.7 \times 10^{-4}$ & -0.3 & ... \\
$B(2^1S_0) \pi$ & -2.8 & 2.7 & ... & ... & ... & ... & -3.9 & 0.2 \\
$B(2^1S_0) \eta$ & -0.2 & ... & ... & ... & ... & ... & ... & ... \\
$B(2^3S_1) \pi$ & -1.8 & 1.5 & -3.4 & 1.2 & -9.7 & 1.5 & -4.2 & 0.2 \\
$B(2^3S_1) \eta$ & -0.1 & ... & ... & ... & -1.0 & ... & ... & ... \\
$B(1^3P_0) \pi$ & ... & ... & -0.2 & $1.7 \times 10^{-2}$ & -4.3 & 12.4 & ... & ... \\
$B_s(1^3P_0) K$ & ... & ... & $-4.2 \times 10^{-2}$ & ... & -1.2 & 0.2 & ... & ... \\
$B(1^3P_0) \eta$ & ... & ... & $-2.8 \times 10^{-2}$ & $5.7 \times 10^{-5}$ & -0.7 & 0.4 & ... & ... \\
$B(1P_1) \pi$ & -1.2 & $1.0 \times 10^{-2}$ & -0.7 & 0.1 & -3.0 & 6.7 & -7.0 & 8.8 \\
$B_s(1P_1) K$ & -0.3 & ... & -0.2 & ... & -0.7 & $7.3 \times 10^{-3}$ & -1.5 & ... \\
$B(1P_1) \eta$ & -0.2 & $4.2 \times 10^{-4}$ & -0.1 & ... & -0.4 & 0.1 & -0.7 & $6.7 \times 10^{-4}$ \\
$B_1(5721) \pi$ & -24.5 & 1.8 & -0.9 & $1.4 \times 10^{-2}$ & -9.9 & 4.9 & -7.2 & 0.5 \\
$B_{s1}(5830) K$ & -4.6 & ... & ... & ... & -2.1 & ... & ... & ... \\
$B_1(5721) \eta$ & -2.1 & ... & -0.1 & ... & -1.0 & $2.0 \times 10^{-3}$ & -0.8 & ... \\
$B_2^*(5747) \pi$ & -6.4 & 0.9 & -23.0 & 1.8 & -15.3 & 2.8 & -12.3 & 2.3 \\
$B_{s2}^*(5840) K$ & -1.3 & ... & ... & ... & -3.5 & ... & ... & ... \\
$B_2^*(5747) \eta$ & -0.6 & ... & -1.8 & ... & -1.7 & $9.5 \times 10^{-3}$ & -1.3 & ... \\
$B(1^3D_1) \pi$ & -1.1 & 0.2 & -1.7 & $1.1 \times 10^{-2}$ & -4.3 & 1.5 & -0.1 & $2.9 \times 10^{-4}$ \\
$B(1D_2) \pi$ & -5.2 & 1.9 & -2.6 & ... & -2.2 & $4.5 \times 10^{-2}$ & -4.5 & $7.5 \times 10^{-3}$ \\
$B(1D'_2) \pi$ & -67.4 & ... & -1.9 & ... & -8.0 & $1.5 \times 10^{-4}$ & -6.2 & ... \\
$B(1^3D_3) \pi$ & -7.5 & $2.3 \times 10^{-2}$ & -74.3 & 22.5 & -12.2 & 3.7 & -10.8 & $4.0 \times 10^{-3}$ \\
$B(2^3P_0) \pi$ & ... & ... & -2.4 & $7.8 \times 10^{-3}$ & -1.9 & 1.2 & ... & ... \\
$B(2P_1) \pi$ & -10.9 & 4.6 & -1.7 & $2.5 \times 10^{-6}$ & -1.4 & $2.6 \times 10^{-2}$ & -3.7 & 0.1 \\
$B(2P_1^\prime) \pi$ & -15.2 & 3.7 & -1.2 & ... & -2.8 & 0.1 & -2.6 & $8.2 \times 10^{-5}$ \\
$B(2^3P_2) \pi$ & -3.1 & ... & -14.6 & ... & -4.4 & 0.1 & -2.4 & ... \\
$B \rho$ & -0.1 & $2.3 \times 10^{-2}$ & -0.4 & 0.3 & $-1.4 \times 10^{-2}$ & 0.4 & -0.1 & $2.5 \times 10^{-2}$ \\
$B \omega$ & $-1.7 \times 10^{-2}$ & $5.9 \times 10^{-3}$ & -0.1 & 0.1 & $-7.1 \times 10^{-3}$ & 0.1 & $-2.0 \times 10^{-2}$ & $6.3 \times 10^{-3}$ \\
$B_s K^*$ & $-9.2 \times 10^{-3}$ & ... & -0.1 & ... & $-3.1 \times 10^{-2}$ & $1.7 \times 10^{-2}$ & $-1.1 \times 10^{-2}$ & ... \\
$B^* \rho$ & -1.2 & 1.2 & -0.5 & 0.1 & -0.3 & 0.8 & -0.3 & 0.5 \\
$B^* \omega$ & -0.4 & 0.3 & -0.2 & $1.9 \times 10^{-2}$ & -0.1 & 0.3 & -0.1 & 0.1 \\
$B_s^* K^*$ & -0.2 & ... & ... & ... & -0.1 & $6.7 \times 10^{-6}$ & ... & ... \\
$B(1^3P_0) \rho$ & -1.8 & ... & ... & ... & -0.2 & ... & ... & ... \\
$B(1^3P_0) \omega$ & -0.6 & ... & ... & ... & -0.1 & ... & ... & ... \\
\hline
$\mathrm{Total}$ & -162.1 & 28.4 & -133.8 & 31.7 & -106.5 & 59.8 & -81.9 & 27.9 \\
\hline\hline
\end{tabular*}
\end{table*}

\begin{table*}[t!]
\caption{Mass shifts $\Delta M_i$ and OZI-allowed partial widths $\Gamma_i$ for the $B_s(1F)$ states. Notation and units are the same as in Table~\ref{tab:B_3S_channel_ccef}.}
\label{tab:Bs_1F_channel_ccef}
\setlength{\tabcolsep}{0.85pt}
\renewcommand{\arraystretch}{1.05}
\begin{tabular*}{\textwidth}{@{\extracolsep{\fill}}lcccccccc@{}}
\hline\hline
  & \multicolumn{2}{c}{$B_s(1^3F_2)$} & \multicolumn{2}{c}{$B_s(1F_3)$} & \multicolumn{2}{c}{$B_s(1F'_3)$} & \multicolumn{2}{c}{$B_s(1^3F_4)$} \\
\cline{2-3}\cline{4-5}\cline{6-7}\cline{8-9}
Channel & $\Delta M_i [6449]$ & $\Gamma_i [6408]$ & $\Delta M_i [6368]$ & $\Gamma_i [6337]$ & $\Delta M_i [6468]$ & $\Gamma_i [6407]$ & $\Delta M_i [6378]$ & $\Gamma_i [6333]$ \\
\hline
$B K$ & $-1.8 \times 10^{-2}$ & 3.7 & ... & ... & ... & ... & -3.5 & 6.6 \\
$B_s \eta$ & $-5.9 \times 10^{-2}$ & 0.2 & ... & ... & ... & ... & -0.4 & 0.4 \\
$B_s \eta'$ & -0.1 & $2.9 \times 10^{-2}$ & ... & ... & ... & ... & -0.3 & $2.3 \times 10^{-7}$ \\
$B^* K$ & $-5.3 \times 10^{-2}$ & 1.9 & -0.5 & 2.8 & -9.3 & 20.5 & -5.1 & 7.7 \\
$B_s^* \eta$ & $-5.3 \times 10^{-2}$ & $5.2 \times 10^{-2}$ & -0.1 & $5.8 \times 10^{-2}$ & -1.1 & 1.2 & -0.5 & 0.4 \\
$B_s^* \eta'$ & -0.1 & $6.0 \times 10^{-3}$ & ... & ... & -1.0 & $1.0 \times 10^{-3}$ & -0.5 & ... \\
$B(2^1S_0) K$ & -2.4 & 2.1 & ... & ... & ... & ... & -3.4 & $4.0 \times 10^{-3}$ \\
$B(2^3S_1) K$ & -1.6 & 0.8 & -2.9 & 0.1 & -8.3 & 0.1 & -3.8 & $7.6 \times 10^{-5}$ \\
$B(1^3P_0) K$ & ... & ... & -0.3 & $2.4 \times 10^{-2}$ & -5.6 & 12.7 & ... & ... \\
$B_s(1^3P_0) \eta$ & ... & ... & $-2.6 \times 10^{-2}$ & $1.0 \times 10^{-3}$ & -0.7 & 0.3 & ... & ... \\
$B(1P_1) K$ & -1.8 & $4.1 \times 10^{-2}$ & -1.1 & 0.1 & -3.7 & 5.5 & -7.8 & 5.0 \\
$B_s(1P_1) \eta$ & -0.2 & $1.0 \times 10^{-3}$ & -0.1 & $5.0 \times 10^{-6}$ & -0.4 & $4.2 \times 10^{-2}$ & -0.8 & $3.6 \times 10^{-4}$ \\
$B_1(5721) K$ & -21.3 & 18.6 & -0.9 & $3.2 \times 10^{-2}$ & -9.6 & 2.9 & -6.5 & 0.1 \\
$B_{s1}(5830) \eta$ & -2.7 & 0.5 & ... & ... & -1.0 & $1.0 \times 10^{-3}$ & -0.7 & ... \\
$B_2^*(5747) K$ & -5.9 & 3.3 & -22.9 & 16.1 & -13.7 & 1.7 & -11.7 & 0.5 \\
$B_{s2}^*(5840) \eta$ & -0.7 & $4.4 \times 10^{-2}$ & ... & ... & -1.5 & $8.0 \times 10^{-3}$ & ... & ... \\
$B(1^3D_1) K$ & -2.1 & ... & ... & ... & -3.5 & ... & ... & ... \\
$B K^*$ & -0.1 & 0.1 & -0.6 & 1.1 & $-2.9 \times 10^{-2}$ & 0.5 & -0.1 & $3.6 \times 10^{-2}$ \\
$B_s \phi$ & $-1.2 \times 10^{-2}$ & $2.0 \times 10^{-6}$ & ... & ... & $-2.6 \times 10^{-2}$ & $3.0 \times 10^{-3}$ & ... & ... \\
$B^* K^*$ & -1.8 & 3.4 & -0.8 & 0.6 & -0.5 & 1.0 & -0.4 & 0.7 \\
$B_s^* \phi$ & -0.3 & ... & ... & ... & $-7.7 \times 10^{-2}$ & ... & ... & ... \\
$B(1^3P_0) K^*$ & ... & ... & ... & ... & -0.2 & ... & ... & ... \\
\hline
$\mathrm{Total}$ & -41.3 & 34.6 & -30.4 & 20.9 & -60.1 & 46.5 & -45.4 & 21.4 \\
\hline\hline
\end{tabular*}
\end{table*}

\bibliography{ref}

\end{document}